\documentclass[journal,draftcls,onecolumn,12pt,onecolumn]{IEEEtranTCOM}
\normalsize

\usepackage{cite}

\usepackage{bm}
\usepackage{amsthm}
\newtheorem{defn}{Definition} 
\usepackage{amssymb}
\newtheorem{theorem}{Theorem}
\usepackage{algorithmic}
\usepackage{caption}
\usepackage{subcaption}

\usepackage[pdftex]{graphicx}
\usepackage[cmex10]{amsmath}
\begin{document}
%
\title{Estimation Theory Based Robust Phase Offset Estimation in the Presence of Delay Attacks}
%
%
%

\author{Anantha K. Karthik,~\IEEEmembership{Student Member,~IEEE,}
        and~Rick~S. Blum,~\IEEEmembership{Fellow,~IEEE}
\thanks{This work was supported in part by the U.S. Army Research Laboratory and the U. S. Army Research office under grant number W911NF-14-1-0245 and by Pennsylvania Infrastructure Technology Alliance (PITA), a partnership of Carnegie Mellon University, Lehigh University, Bethlehem, PA, USA, and the Commonwealth of Pennsylvania, Department of Economic and Community Development (DCED).}
\thanks{A. K. Karthik and R. S. Blum are with the Department of Electrical and Computer Engineering, Lehigh University, Bethlehem, PA 18015 USA (e-mail: akk314\@lehigh.edu; rblum@lehigh.edu).}}

%
%

\markboth{IEEE Transactions on Communications}%
{Submitted paper}
%



\maketitle

\begin{abstract}
This paper addresses the problem of robust clock phase offset estimation for the IEEE 1588 precision time protocol (PTP) in the presence of delay attacks. Delay attacks are one of the most effective cyber attacks in PTP, as they cannot be mitigated using typical security measures. In this paper, we consider the case where the slave node can exchange synchronization messages with multiple master nodes synchronized to the same clock. We first provide lower bounds on the best achievable performance for any phase offset estimation scheme in the presence of delay attacks. We then present a novel phase offset estimation scheme that employs the Expectation-Maximization algorithm for detecting which of the master-slave communication links have been subject to delay attacks. After discarding information from the links identified as attacked, which we show to be optimal, the optimal vector location parameter estimator is employed to estimate the phase offset of the slave node. Simulation results are presented to show that the proposed phase offset estimation scheme exhibits performance close to the lower bounds in a wide variety of scenarios.  
\end{abstract}


%
\IEEEpeerreviewmaketitle

\section{Introduction}\label{section1}
The ability to precisely synchronize clocks among distributed components is critical in many vital fields such as electrical power systems, industrial automation, and telecommunications. The IEEE 1588 ``Precision Time Protocol'' (PTP) \cite{IEEE1588} is a widely used clock synchronization protocol for networked measurement and control systems. Although existing protocols such as the Network Time Protocol (NTP) \cite{NTP} and the Global Positioning System (GPS) are available for synchronizing clocks in some cases, PTP offers an inexpensive alternative achieving sub-microsecond precision in meaningful scenarios. NTP restricts the underlying network to be IP-based, while GPS requires line-of-sight. PTP does not have these restrictions and is often far more cost effective in many applications. In fact, packet-based synchronization techniques based on PTP are a frequently employed alternative for achieving synchronization in many applications. For example, PTP is extremely popular in powerline networks \cite{Gaderer_2005}, and in backhaul networks employed for connecting  cell towers in 4G Long Term Evolution (LTE) mobile networks \cite{Hadzic_2011}. 

In PTP, as with any two-way message exchange schemes, the slave node exchanges a series of synchronization packets with its master node so the packet timestamps can be employed to estimate the phase offset relative to the master node. The synchronization packets can encounter several intermediate switches and routers along the network path between the master and the slave node. The main factors contributing to the overall end-to-end delay for a synchronization packet are the constant propagation delay over the entire path, the processing delays at the intermediate nodes, and the random queuing delays at each such node. The problem of estimating the phase offset of the slave node in the presence of random queuing delays is referred to as the phase offset estimation (POE) problem \cite{Anand_2015}. 

Due to their widespread use,  synchronization protocols such as PTP could become the target for attackers trying to disrupt a network. One of the most effective cyber attacks threatening PTP is the delay attack \cite{Ullmann_2009}. In a delay attack, a malicious intermediate node deliberately delays the transmission of synchronization messages in order to alter the estimated phase offset at the slave node. In this paper, we consider developing robust POE schemes to combat delay attacks.

Garderer et. al \cite{Gaderer_2006}~--\cite{Gaderer_2010} studied  various security aspects of clock synchronization in PTP. They proposed an idea of using a group of masters rather than a single master for synchronizing the slave node. The group of masters coordinate in a first stage of their approach. Afterwards, a speaker node representing the group exchanges timing messages with the slave node. The proposed POE scheme works in the presence of a master failure, or a malicious master, however it requires prior information regarding the number of master nodes that have been attacked or failed. This information might not be  available in many scenarios.

Sun et. al \cite{Sun_2006} proposed a robust POE scheme that uses the median of the observed timing offsets from different master nodes to estimate the phase offset. The proposed POE scheme is robust to attacks, however there is a loss in performance due to the significant amount of information being discarded. Song et. al \cite{Song_2005} proposed a robust POE scheme that employs the {Grubb outliers test} to identify the observations that have been attacked. After discarding the observations identified as attacked, the phase offset is estimated from the remaining observations using the sample mean estimator. 

Mizrahi \cite{Mizrahi_2012}~--\cite{Mizrahi_2012_2} proposed a POE scheme that uses multiple master-slave communication links to improve the accuracy of the phase offset estimate. They also discussed using multiple master-slave communication links to help protect against delay attacks. In the proposed POE scheme, the slave node initially estimates the phase offset corresponding to each master-slave communication link. A master-slave communication link is identified as attacked, if the difference between its phase offset estimate and the average of the remaining phase offset estimates is greater than a certain threshold. The information from links identified as attacked are discarded and the phase offset is estimated using information from the unattacked links. In the above POE schemes \cite{Gaderer_2006}~--\cite{Mizrahi_2012_2}, the information from multiple unattacked sources are fused using the sample mean estimator. Though asymptotically consistent, the sample mean estimator might not be the optimal approach for fusing information in many scenarios with a finite number of observations.

Building on the work of \cite{Mizrahi_2012}~--\cite{Mizrahi_2012_2}, in the presented work, we consider the case where the slave node communicates with multiple master nodes whose clocks are already synchronized. We assume that fewer than half of the master-slave links are attacked\footnote{Either this or some information must be provided to allow the estimator to distinugush between sets of unattacked and sets of similarly attacked links.}. Under the stated assumptions, we first derive a lower bound on the best achievable performance for any POE scheme in the presence of delay attacks by extending the work done in \cite{Anand_2015}. We then present a novel robust POE scheme to combat delay attacks in LTE backhaul networks.

Under these assumptions, we present a POE scheme that does not require prior information regarding the number of master-slave communication links that are attacked and employs the optimal approach for fusing information from unattacked multiple master nodes if the probability density functions (pdf) are estimated correctly. In backhaul networks, the background traffic from other users in the network often results in random queuing delays for the synchronization messages. Based on previous simulation studies \cite{Fang_1999}~--\cite{Tijms_2003},  we model the density function of the random queuing delays as a mixture of Gamma random variables, a very general model. Using the Expectation-Maximization (EM) algorithm, our POE approach will first determine the master-slave communication links which are compromised due to delay attacks. At the same time, the EM algorithm will estimate the mixture coefficients describing the pdf of the random queuing delays \cite{Zhang_2015}. Then motivated by \cite{Zhang_attacks_2015}, we only use the information from links determined to be unattacked to estimate the phase offset. We employ the optimal estimator for fusing information from the unattacked links. We carry out extensive simulations to examine the performance of the proposed POE scheme under various scenarios in the presence of delay attacks. Simulation results show that the proposed POE scheme performs better than the conventional POE schemes available in the literature and exhibits a mean square estimation error close to the lower bounds in a wide variety of scenarios. 

The rest of the paper is organized as follows. In Section \ref{section2}, we briefly discuss PTP and the considered problem statement. In Section \ref{section3}, we describe the EM-algorithm used for identifying the attacked links and the mixture components, along with the optimal approach for fusing information from multiple master nodes which are unattacked and the pdfs are all estimated correctly. In Section \ref{section4}, we give some numerical results highlighting the performance of the proposed approach. Finally in Section \ref{section5}, we present our conclusions.

\section{Problem formulation}\label{section2}
In this section, we briefly describe the two-way message exchange mechanism employed in PTP, along with the considered problem statement. We assume the slave node can exchange synchronization messages with $N$ master nodes synchronized to the same clock. Assuming a total of $P$ rounds of two-way message exchanges between the slave and each master node; the following sequence of messages are exchanged between the $i^{th}$ master node and the slave during the $j^{th}$ round of message exchange. The master node initiates the exchange by sending a \emph{sync} packet to the slave at time $t_{1ij}$. The value of $t_{1ij}$ is later communicated to the slave via a \emph{follow\_up} message. The slave node records the time of reception of the \emph{sync} message as $t_{2ij}$ and sends a \emph{delay\_req} message to the master at time $t_{3ij}$. The master records the time of arrival of the \emph{delay\_req} packet at time $t_{4ij}$ and this value is later communicated to the slave using a \emph{delay\_resp} packet. This procedure can be mathematically modeled as \cite{Anand_2015}, \cite{Wu_2011}--\cite{Anand_Lest}
\begin{eqnarray}
u_{ij} & = & d_i + \delta + w_{1ij}, \label{u_data} \\
v_{ij} & = & d_i - \delta + w_{2ij} \label{v_data},
\end{eqnarray}
where $u_{ij} = \left(t_{2ij} - t_{1ij}\right)$ and $v_{ij} = \left(t_{4ij} - t_{3ij}\right)$ in (\ref{u_data}) and (\ref{v_data}). $\delta$ denotes the unknown phase offset between the master and slave node clocks, $d_i$ is the unknown fixed delay at the $i^{th}$ master-slave communication link, $\{w_{1ij}, w_{2ij}\}$ represent the random queuing delays in the forward- and reverse-path respectively. The random queuing delays $\{w_{kij}\}_{j=1}^P$ have a pdf given by $f_{ki}(w)$ for $i = 1, 2, \cdots, N$ and $k = 1, 2$. 
To launch a delay attack at the $i^{th}$ master-slave link, a malicious node adds a finite amount of constant delay to either the forward or reverse path, so we have
\begin{eqnarray}
u_{ij} & = & \left(d_i + \delta + w_{1ij} + \tau_i\right), \\
v_{ij} & = & \left(d_i - \delta + w_{2ij}\right), 
\end{eqnarray}
where the variable $\tau_i$ represents the magnitude of the delay attack\footnote{If the master-slave link is not attacked, then $\tau_i = 0$.}. When the $i^{th}$ master-slave link is not attacked, the forward-path delay $u_{ij}$ has a pdf given by $f_{u_{ij}}(u_{ij}|d_i, \delta) = f_{1i}(u_{ij} - d_i - \delta)$ which depends on the unknown parameters $d_i$ and $\delta$. When the $i^{th}$ master-slave link is attacked, $u_{ij}$ has a pdf given by $f_{u_{ij}}(u_{ij}|d_i, \delta, \tau_i) = f_{1i}(u_{ij} - d_i - \delta - \tau_i)$. Similarly, the reverse-path delay $v_{ij}$ has a pdf given by $f_{v_{ij}}(v_{ij}|d_i, \delta) = f_{2i}(v_{ij} - d_i + \delta)$ depending on the unknown parameters $d_i$ and $\delta$. In this paper, we model the random queuing delays in the backhaul networks as a mixture of Gamma random variables to allow a very general representation which fits data from realistic simulations well \cite{Fang_1999}~--\cite{Tijms_2003}. This approximation is useful in scenarios where we do not have complete knowledge of the queuing delay distributions and we want to estimate with a model with only a few parameters.
So we have
\begin{eqnarray} \label{queuing_delay}
f_{1i}(w) & = & \sum_{k=1}^{M_i} \alpha_{ik} h_{a_{1ik}, b_{1ik}}(w),  \label{queuing_delay1} \\
f_{2i}(w) & = & \sum_{l=1}^{L_i} \beta_{il} h_{a_{2il}, b_{2il}}(w)\label{queuing_delay2}
\end{eqnarray}
for $i = 1, 2, \cdots, N$. In (\ref{queuing_delay}) and (\ref{queuing_delay2}), $M_i$ and $L_i$ represent the number of mixture components in the forward- and reverse path of the $i^{th}$ master-slave link respectively. The variables $\{\alpha_{ik}\}_{k=1}^{M_i}$ and $\{\beta_{il}\}_{l=1}^{L_i}$ represent the unknown mixing coefficients at the forward- and reverse path of the $i^{th}$ master-slave link respectively with $\sum_{k=1}^{M_i}\alpha_{ik} = 1$ and $\sum_{l=1}^{L_i}\beta_{il} = 1$ and the variables $\{a_{1ik}, b_{1ik}\}_{k=1}^{M_i}$, $\{a_{2il}, b_{2il}\}_{l=1}^{L_i}$ represent the corresponding unknown  parameters of the Gamma distributions. Let $\bm{\Theta}$ denote the vector containing the desired parameter to be estimated $\delta$, the set of unknown fixed delays $\{d_i\}_{i=1}^N$, the set of unknown delay attack magnitudes for the master-slave links $\{\tau_i\}_{i=1}^N$, the set of unknown mixing coefficients  $\{\alpha_{i1}, \cdots, \alpha_{iM_i}\}_{i=1}^N$, $\{\beta_{i1}, \cdots, \beta_{iL_i}\}_{i=1}^N$, the set of unknown Gamma distributions parameters $\{a_{1i1}, \cdots, a_{1iM_i}\}_{i=1}^N$, $\{b_{1i1}, \cdots, b_{1iM_i}\}_{i=1}^N$, $\{a_{2i1}, \cdots, a_{2iL_i}\}_{i=1}^N$, $\{b_{2i1}, \cdots, b_{2iL_i}\}_{i=1}^N$, and the set of unknown binary state variables $\{\eta_i\}$ defined shortly. We have
\begin{eqnarray}
\bm{\Theta} & = & [\bm{\Psi}^{\textrm{T}}, \bm{\alpha}^{\textrm{T}}, \bm{\beta}^{\textrm{T}},   \bm{\eta}^{\textrm{T}}]^\textrm{T},
\end{eqnarray}
where
\begin{eqnarray}
\bm{\Psi} & = & \left[\delta, d_1, \cdots, d_N, \tau_1, \cdots, \tau_N, \bm{a_1}^{\textrm{T}}, \bm{b_1}^{\textrm{T}}, \bm{a_2}^{\textrm{T}}, \bm{b_2}^{\textrm{T}}\right]^{\textrm{T}}, \\
\bm{a_1} & = & \left[a_{111}, \cdots, a_{11M_1}, a_{121}, \cdots, a_{12M_2}, \cdots, a_{1NM_N}\right]^{\textrm{T}}, \\
\bm{b_1} & = & \left[b_{111}, \cdots, b_{11M_1}, b_{121}, \cdots, b_{12M_2}, \cdots, b_{1NM_N}\right]^{\textrm{T}}, \\
\bm{a_2} & = & \left[a_{211}, \cdots, a_{21L_1}, a_{221}, \cdots, a_{22L_2}, \cdots, a_{2NL_N}\right]^{\textrm{T}}, \\
\bm{b_2} & = & \left[b_{211}, \cdots, b_{21L_1}, b_{221}, \cdots, b_{22L_2}, \cdots, b_{2NL_N}\right]^{\textrm{T}},  \\
\bm{\alpha} & = & \left[\alpha_{11}, \cdots, \alpha_{1M_1}, \alpha_{21}, \cdots, \alpha_{2M_2}, \cdots, \alpha_{NM_N}\right]^{\textrm{T}}, \\
\bm{\beta} & = & \left[\beta_{11}, \cdots, \beta_{1L_1}, \beta_{21}, \cdots, \beta_{2L_2}, \cdots, \beta_{NL_N}\right]^{\textrm{T}}, \\
\bm{\eta} & = & [\eta_1, \eta_2, \cdots, \eta_{N}]^{\textrm{T}}.
\end{eqnarray}
The $i^{th}$ element of $\bm{\eta}$ is $1$ if the $i^{th}$ master-slave link is attacked, otherwise it has a value $0$ indicating that the link is not attacked. Let ${\bf u}$ and ${\bf v}$ denote the vectors of all the forward and reverse observations at the slave node. We can represent them by
\begin{eqnarray}
{\bf u} & = & [u_{11}, \cdots, u_{1P}, u_{21}, \cdots, u_{NP}]^{\textrm{T}}, \\
{\bf v} & = & [v_{11}, \cdots, v_{1P}, v_{21}, \cdots, v_{NP}]^{\textrm{T}}, 
\end{eqnarray}
and
\begin{eqnarray}
{\bf y} & = & [{\bf u}^{\textrm{T}}, {\bf v}^{\textrm{T}}]^{\textrm{T}}. \label{Fulldata} 
\end{eqnarray}
In our work, we seek an estimator $\hat{\delta}({\bf y})$ of the phase offset $\delta$ based on the observations ${\bf y}$ which are generally effected by all the deterministic parameters in $\bm{\Theta}$. We characterize the performance of $\hat{\delta}({\bf y})$ via the conditional mean squared error (MSE) metric defined as
\begin{eqnarray}\label{mse}
\mathcal{R}\left(\hat{\delta}({\bf y}), \delta\right) & = & E\left\{\left[\delta - \hat{\delta}({\bf y})\right]^2\middle|\bm{\Theta}\right\},
\end{eqnarray}
where $E\left\{.\right\}$ denotes the expectation operator.

\section{Attack identification and data fusion}\label{section3}
In this section, we develop our robust approach. First, we present the optimum scheme for fusing information exchanged with the various master nodes when we know which master-slave communication links are attacked. The approach employs our previous results on unattacked links and on when to use data from attacked links in general estimation problems. To pull together these ideas, a modified EM algorithm for identifying the links which have been attacked incorporates our developed theory on fusing the attacked and unattacked data. We first present an important theorem, which we use for the remainder of this paper.

\begin{theorem}\label{thm1}
There is no gain in estimating $\delta$ using the information from the attacked master-slave communication links (For proof: See Theorem 1 of \cite{Zhang_attacks_2015}). 
\end{theorem}
As a consequence of Theorem \ref{thm1}, we see that discarding the observations from the attacked links does not affect the performance of an estimator of the phase offset $\delta$.

\subsection{Genie approach}
Let's assume that we have \emph{prior} knowledge of the master-slave links which have been attacked, as well as complete knowledge of the forward- and reverse queuing delay pdfs $f_{1i}(w)$ and $f_{2i}(w)$ for all $N$ master-slave links $i = 1, 2, \cdots, N$. Following Theorem \ref{thm1}, we discard the observations from the attacked links. We employ the unattacked minimax optimum estimator presented in \cite{Anand_2015} to present an  approach which can provide a very useful but generally unachievable genie lower bound on the performance of estimating $\delta$ in our general problem. We first present some definitions and results from \cite{Anand_2015}. 

\begin{defn} (Vector location Parameter problem): Suppose we want to estimate a linear combination ${\bf c}^T\bm{\theta}$ of the unknown parameters contained in the components of the vector $\bm{\theta} \in \mathbb{R}^M$ (where ${\bf c} \in \mathbb{R}^M$ is a constant vector), based on observations ${\bf x} \in \mathbb{R}^N$.  If the observations have a pdf of the form
	\begin{eqnarray}\label{vec_loc_parameter}
	f({\bf x}|\bm{\theta}) = f_0({\bf x} - {\bf G}\bm{\theta})
	\end{eqnarray}
	for some $N \times M$ matrix ${\bf G}$, and function $f_0(.)$, such a problem is referred to as a {vector location parameter problem} \cite{Anand_2015}.
\end{defn}

\begin{defn}
	We say that an estimator $g({\bf x})$ of ${\bf c}^T\bm{\theta}$ is shift invariant if for the same matrix ${\bf G}$ used in (\ref{vec_loc_parameter}),
	\begin{eqnarray}
	g({\bf x + Gh}) & = & g({\bf x}) + {\bf c}^T{\bf h}
	\end{eqnarray}
	for all ${\bf h} \in \mathbb{R}^M$ \cite{Anand_2015}.
\end{defn}

\noindent The conditional MSE of an estimator $g({\bf x})$ of ${\bf c}^T\bm{\theta}$ can be written as
\begin{eqnarray}
\mathcal{R}(g({\bf x}), {\bf c}^T\bm{\theta}) & = & \int_{\mathbb{R}^N}[g({\bf x}) - {\bf c}^T\bm{\theta}]^2f({\bf x}|\bm{\theta})d{\bf x}
\end{eqnarray}
and the maximum MSE is defined as
\begin{eqnarray}
\mathcal{M}(g({\bf x})) & = & \sup_{\bm{\theta} \in \mathbb{R}^M}\mathcal{R}(g({\bf x}), {\bf c}^T\bm{\theta}).
\end{eqnarray}
While the conditional and maximum MSE are in general different for an estimator, it was shown in \cite{Anand_2015} that for a shift invariant estimator they are equal. Guruswamy et. al \cite{Anand_2015} derived a useful optimum estimator for the vector location parameter estimation problem without any attacks. The optimum estimator is presented below for convenience.
\begin{theorem}\label{thm2}
	For the vector location parameter problem, the estimator
	\begin{eqnarray}\label{opt_estimator}
	g^*({\bf x}) & = & \frac{\int_{\mathbb{R}^M}[{\bf c}^T\bm{\theta}]f({\bf x}|\bm{\theta})d\bm{\theta}}{\int_{\mathbb{R}^M}f({\bf x}|\bm{\theta})d\bm{\theta}},
	\end{eqnarray}
	satisfies the following properties:
	\begin{itemize}
		\item $g^*({\bf x})$ is shift invariant. In fact, all the estimators used in practice for POE are shift invariant.
		\item $g^*({\bf x})$ minimizes the maximum MSE for any estimate ${\bf c}^T\bm{\theta}$.
		\item Among all estimators of ${\bf c}^T\bm{\theta}$ that are shift invariant, $g^*({\bf x})$ achieves the minimum conditional MSE.
		\item $g^*({\bf x})$ is unbiased, i.e., $E\left\{\left[g^*({\bf x}) - {\bf c}^T\bm{\theta}\right] | \bm{\theta}\right\} = 0$.
	\end{itemize}
	For proof, see \cite{Anand_2015}.
\end{theorem}
\noindent The estimator $g^*({\bf x})$ presented in (\ref{opt_estimator}) of ${\bf c}^T\bm{\theta}$ is optimum in terms of minimizing the conditional mean-squared estimation error over all values of the unknown parameters among shift invariant estimators. In this paper, we use the optimum vector location parameter estimator presented in Theorem \ref{thm2} along with the result from Theorem \ref{thm1} to get a lower bound on the MSE for a POE scheme in the presence of delay attacks.

If we have \emph{prior} information regarding the links which have been attacked, Theorem 1 tells us to only retain information from the unattacked links. Without loss of generality, let us assume that the $\textrm{K}$ master-slave links indexed by $\{n_k\}_{k=1}^K$ are unattacked,  the observed data from these links can be represented as
\begin{eqnarray}
{\bf u}_K & = & [u_{n_11}, \cdots, u_{n_1P}, u_{n_21}, \cdots, u_{n_KP}]^{\textrm{T}}, \\
{\bf v}_K & = & [v_{n_11}, \cdots, v_{n_1P}, v_{n_21}, \cdots, v_{n_KP}]^{\textrm{T}}, \\
{\bf y}_K & = & [{\bf u}_K^{\textrm{T}}, {\bf v}_K^{\textrm{T}}]^{\textrm{T}}.
\end{eqnarray}
From (3) and (4), we can represent ${\bf y}_K$ as
\begin{eqnarray}\label{genie_data}
{\bf y}_K & = & {\bf G}\boldsymbol{\theta} + {\bf w}_K,
\end{eqnarray}
where $\boldsymbol{\theta} = [\delta, d_{n_1}, \cdots, d_{n_K}]^{\textrm{T}}$ represents the vector of unknown parameters and ${\bf w}_K$ is a random $\textrm{KP} \times 1$ vector given by ${\bf w}_K = [w_{1n_11}, w_{1n_12}, \cdots, w_{1n_1P}, w_{1n_21}, \cdots, w_{1n_KP}$, $w_{2n_11}$, $w_{2n_12}$, $\cdots$, $w_{2n_1P}$, $w_{2n_21}, \cdots, w_{2n_KP}]^{\textrm{T}}$. The matrix ${\bf G}$ is given by
\begin{eqnarray}
{\bf G} & = & \begin{bmatrix}
\mbox{ }{\bf 1}_{\textrm{KP} \times 1} & {\bf A}_{\textrm{KP} \times \textrm{K}}  \\
-{\bf 1}_{\textrm{KP} \times 1} & {\bf A}_{\textrm{KP} \times \textrm{K}}                                                                                                                                                                                                                                                                                                             
\end{bmatrix}
\end{eqnarray}
where ${\bf 1}_{\textrm{KP} \times 1}$ represents a vector of size $\textrm{KP} \times 1$ with all the elements equal to $1$, ${\bf A}_{\textrm{KP} \times \textrm{K}} = {\mathbf 1}_{\textrm{P} \times 1} \otimes {\mathbf I}_{\textrm{K} \times \textrm{K}}$, with $\otimes$ representing the Kronecker delta product, and ${\mathbf I}_{\textrm{K} \times \textrm{K}}$ represents an identity matrix of size $\textrm{K}$. We observe that (\ref{genie_data}) has a pdf which can be represented as
\begin{eqnarray}
f_{{\bf y}_K}({\bf y}_K|\boldsymbol{\theta}) = f_{\bf w}({\bf y}_K - {\bf G}\boldsymbol{\theta}).
\end{eqnarray}
Assuming the components of the vector ${\bf w}$ in (\ref{genie_data}) are independent and identically distributed, $f_{\bf w}({\bf w})$  is given by
\begin{eqnarray}
f_{\bf w}({\bf w}) & = & \prod_{i=1}^K\prod_{j=1}^P f_{1n_i}(w_{1n_ij})f_{2n_i}(w_{2n_ij}).
\end{eqnarray}
So our problem of fusing known-attack-status information from multiple master nodes falls under the class of vector location parameter estimation problems. Therefore, we use the optimal estimator given in Theorem \ref{thm2} to obtain the phase offset estimate.

In our problem, we have $\bm{\theta} = [\delta, d_{n_1}, \cdots, d_{n_K}]^{\textrm{T}}$, so we employ ${\bf c} = [1, 0, \cdots 0]^{\textrm{T}}$. By applying Theorem \ref{thm2}, we can then obtain the optimum estimator of $\delta$ from the observations ${\bf y}_K$ as
\begin{eqnarray}\label{optimum_estimator}
\hat{\delta}({\bf y}_K) & = & \frac{\int_{\delta} \delta \Omega(\delta, {\bf u}_K, {\bf v}_K) d\delta}{\int_{\delta} \Omega(\delta, {\bf u}_K, {\bf v}_K)d\delta},
\end{eqnarray}
where $\Omega(\delta, {\bf u}_K, {\bf v}_K)$ is defined as follows
\begin{eqnarray}
\Omega(\delta, {\bf u}_K, {\bf v}_K) & = & \prod_{i = 1}^K\int_{d_{n_i}}\prod_{j=1}^{P}\left[f_{1n_i}(u_{n_ij} - \delta - d_{n_i}) f_{2n_i}(v_{n_ij} + \delta - d_{n_i})\right]d (d_{n_i}). \nonumber
\end{eqnarray}
We refer to the estimator presented in (\ref{optimum_estimator}) as the genie-optimum estimator. The genie optimum estimator gives us a lower bound on the MSE for a POE scheme in the presence of delay attacks. We use this bound to evaluate the performance of our proposed POE scheme.

\subsection{EM-algorithm}
In practice, we do not have \emph{prior} information regarding the links which have been attacked. It is necessary to identify these links before applying the optimal estimator presented in (\ref{optimum_estimator}). In our work, we use the EM algorithm combined with the random relaxation proposed in \cite{Zhang_2015} for identifying these links. For ease of notation, we define $h_{a, b}(w|\theta_1, \theta_2) = h_{a, b}(w - \theta_1 - \theta_2)$, where $\theta_1$ and $\theta_2$ represent unknown deterministic parameters. The log-likelihood function of $\bm{\Theta}$ evaluated using the observed data ${\bf y}$ is given by
\begin{eqnarray}\label{sec3_eq1}
L(\bm{\Theta}|{\bf y}) & = & \sum_{i=1}^{N}\sum_{j=1}^{P}\eta_i \ln \left[\sum_{k=1}^{M_i}\alpha_{ik}h_{a_{1ik}, b_{1ik}}(u_{ij}|\delta, d_i, \tau_i) \sum_{l=1}^{L_i}\beta_{il}h_{a_{2il}, b_{2il}}(v_{ij}|-\delta, d_i)
\right] \nonumber \\
& = & + (1 - \eta_i)\ln \left[\sum_{k=1}^{M_i}\alpha_{ik}h_{a_{1ik}, b_{1ik}}(u_{ij}|\delta, d_i) \sum_{l=1}^{L_i}\beta_{il}h_{a_{2il}, b_{2il}}(v_{ij}|-\delta, d_i)\right].
\end{eqnarray}
Recall $\eta_i$ represents an unknown deterministic binary variable. Based on the power of a post processing (to be discussed later) which is able to undo this modification, we introduce a relaxation that essentially replaces each $\eta_i$ with a real valued variable $\pi_i = \textrm{Pr}(\eta_i = 1) \in [0, 1]$ which represents the probability of the $i^{th}$ link being attacked.  Let $\bm{\Theta}_{\pi} = [\bm{\Psi}^{\textrm{T}}, \bm{\alpha}^{\textrm{T}}, \bm{\beta}^{\textrm{T}}, \bm{\pi}^{\textrm{T}}]^{\textrm{T}}$, where $\bm{\pi} = [\pi_1, \pi_2, \cdots, \pi_N]^{\textrm{T}}$. The log-likelihood function of $\bm{\Theta_{\pi}}$ given the observed vector ${\bf y}$ is given by
\begin{eqnarray}\label{LLF}
L(\bm{\Theta}_{\pi}|{\bf y}) & = & \sum_{i=1}^{N}\sum_{j=1}^{P} \ln \left[\pi_i\sum_{k=1}^{M_i}\sum_{l=1}^{L_i}\alpha_{ik}h_{a_{1ik}, b_{1ik}}(u_{ij}|\delta, d_i, \tau_i) \beta_{il}h_{a_{2il}, b_{2il}}(v_{ij}|-\delta, d_i)
\right. \nonumber \\
& = & + \left.(1 - \pi_i)\sum_{k=1}^{M_i}\sum_{l=1}^{L_i}\alpha_{ik}h_{a_{1ik}, b_{1ik}}(u_{ij}|\delta, d_i) \beta_{il}h_{a_{2il}, b_{2il}}(v_{ij}| -\delta, d_i)\right].
\end{eqnarray}
A maximum-likelihood estimate of $\bm{\Theta}_{\pi}$ can then be obtained using
\begin{eqnarray}\label{Main_Equation2}
\hat{\bf \Theta}_{\pi} & = & \underset{\bf \Theta_{\pi}}{\mbox{arg max }}L(\bm{\Theta}_{\pi}|{\bf y}).
\end{eqnarray}
We use the EM algorithm to evaluate (\ref{Main_Equation2}) to get the maximum likelihood estimate $\hat{\bm{\Theta}}_{\pi}$. We introduce the latent vector ${\bf z}  = [z_{11}, z_{12}, \cdots, z_{1P}, z_{21}, \cdots, z_{NP}]^{\textrm{T}}$, where $z_{ij} = 1$ indicates that the $j^{th}$ two-way timing exchange at the $i^{th}$ master-slave link was attacked, while $z_{ij} = 0$ indicates that it was not attacked, and the latent vector ${\bf r} = [r_{11}, r_{12}, \cdots, r_{1P}, r_{21}, \cdots, r_{NP}]^{\textrm{T}}$ where $r_{ij} = k$ indicates that $u_{ij}$ is from the $k^{th}$ mixture component in the $i^{th}$ forward queuing delay pdf, and $r_{ij} \in \{1, 2, \cdots, M_i\}$. Similarly we introduce the latent vector ${\bf s} = [s_{11}, s_{12}, \cdots, s_{1P}, s_{21}, \cdots, s_{NP}]^{\textrm{T}}$ where $s_{ij} = k$ indicates that $v_{ij}$ is from the $k^{th}$ mixture component in the $i^{th}$ reverse queuing delay pdf, and $s_{ij} \in \{1, 2, \cdots, L_i\}$. The EM algorithm is an iterative approach which alternates between performing an expectation (E) step and a maximization (M) step to calculate the maximum-likelihood estimate of a parameter. The steps of the algorithm are described below:
\begin{enumerate}
\item {\em Q-step}: In this step, we evaluate the expected log-likelihood function $Q(\bm{\Theta}_{\pi}|\bm{\Theta}_{\pi}^g)$ averaged over the unknown data $\{{\bf z}, {\bf r}, {\bf s}\}$, and conditioned on the current estimate of $\bm{\Theta}_{\pi}$ represented by $\bm{\Theta}_{\pi}^g$. We have
\begin{eqnarray}
Q(\bm{\Theta}_{\pi}|\bm{\Theta}_{\pi}^g) & = & E\left\{L\left(\bm{\Theta}_{\pi}|{\bf z, r, s, y})\middle|{\bf y}, \bm{\Theta}^{(g)}\right)\right\},
\end{eqnarray}
where $L(\bm{\Omega}_{\pi}|{\bf z, r, s, y})$ is defined as
\begin{eqnarray}
L(\bm{\Theta}_{\pi}|{\bf z, r, s, y}) & = & \sum_{i=1}^N\sum_{j=1}^{P}\chi_{\{z_{ij}=1\}} \ln \left[\pi_i\alpha_{r_{ij}}\beta_{s_{ij}}h_{a_{1r_{ij}}b_{1r_{ij}}}(u_{ij}|\delta, d_i, \tau_i)h_{a_{2s_{ij}} b_{2s_{ij}}}(v_{ij}| -\delta, d_i)\right] \nonumber \\
& & + \chi_{\{z_{ij}=0\}} \ln \left[(1 - \pi_i)\alpha_{r_{ij}}\beta_{s_{ij}}h_{a_{1r_{ij}} b_{1r_{ij}}}(u_{ij}|\delta,  d_i)h_{a_{2s_{ij}}b_{2s_{ij}} }(v_{ij}|-\delta, d_i)\right], \nonumber\\
\end{eqnarray}
where ${\bf y}$ is defined in (\ref{Fulldata}), and $\chi_{z_{ij}}$ represents the indicator function defined as
\begin{eqnarray}
\chi_{\{z_{ij} = 1\}} = \begin{cases}
1 & \textrm{if } z_{ij} = 1 \\
0 & \textrm{otherwise}.
\end{cases}
\end{eqnarray} The expected log-likelihood function $Q(\bm{\Theta}_{\pi}|\bm{\Theta}_{\pi}^g)$ can then be expressed as
\begin{eqnarray}
Q(\bm{\Theta}_{\pi}|\bm{\Theta}_{\pi}^g) & = & \sum_{i=1}^N\sum_{j=1}^{P}\sum_{k=1}^{M_i}\sum_{l=1}^{L_i}a^{(1)}_{ijkl} \ln \left[\pi_i\alpha_{ik}\beta_{il}h_{a_{1ik}, b_{1ik}}(u_{ij}|\delta, d_i, \tau_i)h_{a_{2il}, b_{2il}}(v_{ij}|-\delta, d_i)\right] \nonumber \\
& & + a^{(0)}_{ijkl} \ln \left[(1 - \pi_i)\alpha_{ik}\beta_{il}h_{a_{1ik}, b_{1ik}}(u_{ij}|\delta, d_i)h_{a_{2il}, b_{2il}}(v_{ij}|-\delta, d_i)\right],  
\end{eqnarray}
where $a^{(1)}_{ijkl}$ and $a^{(0)}_{ijkl}$ are defined as 
\begin{eqnarray}\label{a_ijkl}
a^{(1)}_{ijkl} & = & \textrm{Prob}\{\chi_{\{z_{ij}=1\}} = 1, \alpha_{ik} = r_{ij}, \beta_{il} = s_{ij}|\bm{\Theta}_{\pi}^g\}, \nonumber \\
&  = & \frac{\pi_i^g\alpha_{ik}^g\beta_{il}^gh_{a^g_{1ik}, b^g_{1ik}}(u_{ij}|\delta^g, d_i^g, \tau_i^g)h_{a^g_{2il}, b^g_{2il}}(v_{ij}|-\delta^g, d_i^g)}{\sum_{k_i=1}^{M_i}\sum_{l_i=1}^{L_i}\alpha_{ik_i}^g\beta_{il_i}^gh_{a^g_{2il_i}, b^g_{2il_i}}(v_{ij}|-\delta^g, d_i^g)D_{k_i}}, \nonumber \\
a^{(0)}_{ijkl} & = &  \textrm{Prob}\{\chi_{\{z_{ij}=1\}} = 0, \alpha_{ik} = r_{ij}, \beta_{il} = s_{ij}|\bm{\Theta}_{\pi}^g\}, \nonumber \\
&  = & \frac{(1 - \pi_i^g)\alpha_{ik}^g\beta_{il}^gh_{a^g_{1ik},b^g_{1ik}}(u_{ij}|\delta^g, d_i^g)h_{a^g_{2il}, b^g_{2il}}(v_{ij}|-\delta^g, d_i^g)}{\sum_{k_i=1}^{M_i}\sum_{l_i=1}^{L_i}\alpha_{ik_i}^g\beta_{il_i}^gh_{a^g_{2il_i}, b^g_{2il_i}}(v_{ij}|-\delta^g, d_i^g)D_{k_i}}, \nonumber \\
D_{k_i} & = & \left[\pi_i^gh_{a^g_{1ik_i}, b^g_{1ik_i}}(u_{ij}|\delta^g, d_i^g, \tau_i^g) + (1 - \pi_i^g)h_{a^g_{2ik_i}, b^g_{2ik_i}}(u_{ij}|\delta^g, d_i^g)\right].
\end{eqnarray}

\item {\em M-step}: After we calculate the expected log-likelihood with respect to the unknown data, we update the value of the vector parameter ${\hat{\bm{\Theta}}_{\pi}}$ by maximizing the expected log-likelihood function $Q(\bm{\Theta}_{\pi}|\bm{\Theta}_{\pi}^g)$. We have
\begin{eqnarray}
\hat{\bm{\Theta}}_{\pi} & =  \left[\hat{\bm{\Psi}}^{\textrm{T}},  \hat{\bm{\alpha}}^{\textrm{T}}, \hat{\bm{\beta}}^{\textrm{T}}, \hat{\bm{\pi}}^{\textrm{T}} \right]& = \underset{\bf \Theta_{\pi}}{\mbox{arg max }}Q(\bm{\Theta}_{\pi}|\bm{\Theta}_{\pi}^g).
\end{eqnarray}
We first solve the maximization problem to update $\hat{\pi}_i$, we have
\begin{eqnarray}
\frac{\partial Q(\bm{\Theta}_{\pi}|\bm{\Theta}_{\pi}^g)}{d \pi_i} & = & 0, \\
\sum_{j=1}^P\sum_{k=1}^{M_i}\sum_{l=1}^{L_i}\frac{a^{(1)}_{ijkl}}{\pi_i} - \frac{a^{(0)}_{ijkl}}{1 - \pi_i} & = & 0 .
\end{eqnarray}
Solving the above equation, we get the updated estimate of $\pi_i$ as
\begin{eqnarray}\label{pi_i}
\hat{\pi}_i & = & \frac{1}{P}\sum_{j=1}^P\sum_{k=1}^{M_i}\sum_{l=1}^{L_i}a_{ijkl}^{(1)}.
\end{eqnarray}
To estimate the mixing coefficient $\hat{\alpha}_{ik}$, we introduce the Lagrange multiplier with the constraint $\sum_{k_i=1}^{M_i}\alpha_{ik_i} = 1$, and solve the equation 
\begin{eqnarray}
\frac{\partial }{\partial \alpha_{ik}}\left[Q(\bm{\Theta}_{\pi}|\bm{\Theta}_{\pi}^g) - \lambda\left(\sum_{k_i=1}^{M_i}\alpha_{ik_i} - 1\right)\right] & = & 0.
\end{eqnarray}
We then have
\begin{eqnarray}
\sum_{j=1}^P\sum_{l=1}^{L_i} \frac{1}{\alpha_{ik}}\left(a^{(1)}_{ijkl} + a^{(0)}_{ijkl}\right) - \lambda & = & 0, \\
\sum_{j=1}^P\sum_{l=1}^{L_i} \left(a^{(1)}_{ijkl} + a^{(0)}_{ijkl}\right) & = & \lambda\alpha_{ik}.
\end{eqnarray}
Summing over $k$ on both sides, we can find the value of $\lambda$. We have
\begin{eqnarray}
\lambda\sum_{k=1}^{M_i}\alpha_{ik}   & = & \sum_{k=1}^{M_i}\sum_{j=1}^P\sum_{l=1}^{L_i} \left(a^{(1)}_{ijkl} + a^{(0)}_{ijkl}\right)  \\
\lambda  & = & \sum_{k=1}^{M_i}\sum_{j=1}^P\sum_{l=1}^{L_i} \left(a^{(1)}_{ijkl} + a^{(0)}_{ijkl}\right) \\ 
& = & P.
\end{eqnarray}
So, the updated estimate is given by
\begin{eqnarray}\label{alpha_ik}
\hat{\alpha}_{ik} & = & \frac{1}{P}\sum_{j=1}^P\sum_{l=1}^{L_i} \left(a^{(1)}_{ijkl} + a^{(0)}_{ijkl}\right).
\end{eqnarray}
Following a similar process for $\beta_{il}$, we can get the updated estimate of $\hat{\beta}_{il}$ as
\begin{eqnarray}\label{beta_il}
\hat{\beta}_{il} & = & \frac{1}{P}\sum_{j=1}^P\sum_{k=1}^{M_i} \left(a^{(1)}_{ijkl} + a^{(0)}_{ijkl}\right).
\end{eqnarray}
Calculating the just described updates for all the $N$ master-slave links, we obtain the updated estimate of $\hat{\bm{\alpha}}$ and $\hat{\bm{\beta}}$. An updated estimate of $\hat{\bm{\Psi}}$ can be obtained by solving the equation
\begin{eqnarray}\label{EM_Mstep}
\nabla_{\bm{\Psi}}Q(\bm{\Theta}_{\pi}|\bm{\Theta}_{\pi}^g) & = & {\bf 0}.
\end{eqnarray}
In order to solve (\ref{EM_Mstep}), we use the Newton-Raphson method to get an updated estimate of the parameter $\hat{\bm{\Psi}}$. We choose the initial point as $\bm{\Psi}^{(0)} = \hat{\bm{\Psi}}^g$. We then update the value at the $(t + 1)^{th}$ stage using 
\begin{eqnarray}\label{NewtonRaphson}
\hat{\bm{\Psi}}^{(t+1)} & = & \hat{\bm{\Psi}}^{(t)} - \kappa_{t}\left[\nabla^2_{\bm{\Psi}}Q(\bm{\Theta}^{(t)}_{\pi}|\bm{\Theta}_{\pi}^g)\right]^{-1}\nabla_{\bm{\Psi}}Q(\bm{\Theta}^{(t)}_{\pi}|\bm{\Theta}_{\pi}^g),
\end{eqnarray}
where $\bm{\Theta}^{(t)}_{\pi} = [\hat{\bm{\Psi}}^{\textrm{T}}, \hat{\bm{\alpha}}^{\textrm{T}}, \hat{\bm{\beta}}^{\textrm{T}}, \hat{\bm{\pi}}^{\textrm{T}}]^T$ and $\kappa_{t} \in (0, 1)$  is the $t^{th}$ step-size computed using a backtracking line search \cite{Boyd_2004}. Repeating the calculation in (\ref{NewtonRaphson}) until $\hat{\bm{\Psi}}^{(t)}$ converges, the limit is a solution for (\ref{EM_Mstep}).  
\end{enumerate}
The convergence analysis for the EM algorithm appears in \cite{Wu_1983}. By iteratively alternating between the \emph{E-step} and \emph{M-step}, we  obtain the maximum-likelihood estimate of $\bm{\Theta}_{\pi}$. We then classify the links based on the converged value of $\hat{\bm{\pi}}$. Specifically, we declare the $i^{th}$ master-slave link as attacked if $\pi_i > (1 - \pi_i)$, else we declare the link as unattacked. Following Theorem 1, we discard the observations from the attacked links. We then use the estimated values of $\hat{\bm{a}}_1$, $\hat{\bm{b}}_1$, $\hat{\bm{a}}_2$, $\hat{\bm{b}}_2$, $\hat{\bm{\alpha}}$ and $\hat{\bm{\beta}}$ in the optimal estimator presented in (\ref{optimum_estimator}) to estimate the phase offset $\delta$. As our problem is not necessarily convex, proper initialization of the various parameters is crucial for the EM algorithm to ensure convergence to the global minimum instead of a local minimums. We present a simple ad-hoc approach in Section \ref{EM_init} for initializing the parameters for the EM algorithm. We observe from simulations that the proposed ad-hoc scheme seems to avoid  local minimums. The complete algorithm is summarized in Section \ref{EMAlgo}.

\subsection{Initialization for the EM-algorithm}\label{EM_init}
We now present a simple ad-hoc approach to fix the initial values of $\bm{\Theta}_{\pi}$ denoted by $\bm{\Theta}_{\pi}^{(g)}$ for the EM algorithm described in Section \ref{section3}. We assume prior knowledge regarding the mean and covariance of the forward- and reverse queuing delays for each of the master-slave communication links. Consider the $i^{th}$ master-slave communication link: 
\begin{enumerate}
	\item \emph{Initialization of $\bm{\alpha}^{(g)}$ and $\bm{\beta}^{(g)}$}: We set the initial values of $\left\{\alpha_{ik}^{(g)}\right\}_{k=1}^{M_i}$ as $1/M_i$ and the initial values of $\left\{\beta_{il}^{(g)}\right\}_{l=1}^{L_i}$ as $1/L_i$. This procedure is repeated for all the master-slave communication links in both the forward- and reverse path to get $\bm{\alpha}^{(g)}$ and $\bm{\beta}^{(g)}$.

	\item \emph{Initialization of ${\bm{a}}_1^{(g)}$, ${\bm{b}}_1^{(g)}$, ${\bm{a}}_2^{(g)}$, ${\bm{b}}_2^{(g)}$}: The queuing delay pdf in the forward path is given by (\ref{queuing_delay}). Since we have prior information regarding the mean ($\mu_{1i}$) and variance ($\sigma^2_{1i}$) of the queuing delay distributions, we have
	\begin{eqnarray}
	\mu_{1i} & = & \sum_{k=1}^{M_{i}}\frac{a_{1ik}b_{1ik}}{M_{i}}, \\
	\sigma_{1i}^2 & = & \sum_{k=1}^{M_{i}}\frac{(a_{1ik}b_{1ik}^2 + a_{1ik}^2b_{1ik}^2)}{M_{i}} - a_{1ik}^2b_{1ik}^2.
	\end{eqnarray}
	Solving the above equations, we can get our estimate of $\{a_{1ik}^{(g)}\}_{k=1}^{L_{i}}$ and $\{b_{1il}^{(g)}\}_{l=1}^{L_{i}}$. Following a similar procedure for the reverse path, we obtain  $\{a_{2ik}^{(g)}\}_{k=1}^{M_{i}}$ and $\{b_{2il}^{(g)}\}_{l=1}^{L_{i}}$. This procedure is repeated for all the master-slave communication links in both the forward- and reverse path to get ${\bm{a}}_1^{(g)}$, ${\bm{b}}_1^{(g)}$, ${\bm{a}}_2^{(g)}$, and ${\bm{b}}_2^{(g)}$.
	
	\item \emph{Initialization of $\bm{\pi}$}: We use the POE scheme proposed in \cite{Anand_Lest} in our ad-hoc scheme to initialize $\bm{\pi}^{(g)}$. The POE scheme presented in \cite{Anand_Lest}, referred to as an $L$-estimator is based on a linear combination of the order statistics. These POE schemes offer near optimal performance and only require information regarding the mean and covariance matrix of the queuing delays\footnote{A brief discussion of the POE scheme proposed in \cite{Anand_Lest} is presented in Appendix \ref{Appendix1}.}. The steps to fix $\bm{\pi}^{(g)}$ are as follows:
	\begin{itemize}
		\item We determine the relative phase offset $\hat{\delta}_i$ for all the $N$ master-slave communication links using the \emph{L}-estimators. 
		\item Since less than half of the master-slave links are attacked, the median of these $N$ phase offsets should correspond to a phase offset estimate which is unattacked. We represent this value by $\hat{\delta}_{\bf med}$.
		\item Representing the root mean sqaure error of the phase offset estimate determined by the median as ${\sigma}_{\bf medP}$ (See (\ref{Lest_MSE}) in Appendix). If we observe that $\left|\hat{\delta}_{\bf med} - \hat{\delta}_i\right| \ge 2{\sigma}_{\bf medP}$, we declare the link as attacked, else we declare the link as unattacked.
	\end{itemize}	
	\item \emph{Initialization of $\delta$}: We fix the value of $\delta^{(g)}$ as $\hat{\delta}_{\bf med}$.
	\item \emph{Initialization of $\bm{d}, \bm{\tau}$}: If the $i^{th}$ master-slave link is identified as unattacked, we fix the initial values of $d_i$ and $\tau_i$ as follows:
	\begin{eqnarray}
	d_i^{(g)} & = & \frac{1}{2P}\sum_{j=1}^P\left(u_{ij} + v_{ij}\right) - \frac{(\mu_{1i} + \mu_{2i})}{2}, \\		
	\tau_i^{(g)} & = & 0.
	\end{eqnarray}
	If the $i^{th}$ master-slave link is identified as attacked, we have
	\begin{eqnarray}
	d_i^{(g)} & = & \frac{1}{P}\sum_{j=1}^P v_{ij} + \delta^{(g)} - \mu_{2i}, \\
	\tau_i^{(g)} & = & \frac{1}{P}\sum_{j=1}^P u_{ij} - \delta^{(g)} - d_i^{(g)} - \mu_{1i}.
	\end{eqnarray}	
	This procedure is repeated for all the master-slave communication links to get ${\bm{d}}^{(g)}$ and ${\bm{\tau}}^{(g)}$.
\end{enumerate}

\subsection{Final Algorithm}\label{EMAlgo}
The steps of the proposed POE scheme is presented in this section. We repeat the algorithm until the difference between the complete data log likelihood at consecutive iterations becomes less than some threshold.
\begin{algorithmic}[1]
	\STATE Initialize the parameter vector $\bm{\Theta}_{\pi}$ as $\bm{\Theta}_{\pi}^{(g)}$.
	\STATE Initialize the present value of the log likelihood $\mbox{LLF}_{pres}$ using (\ref{LLF}) and  $\bm{\Theta}_{\pi}^{(g)}$.
	\STATE Initialize the previous value of the log likelihood $\mbox{LLF}_{prev} = -\infty$ and fix a threshold value $\epsilon$ to determine when to stop.
	\WHILE {$\left|\mbox{LLF}_{pres} - \mbox{LLF}_{prev}\right| \ge \epsilon$}
	\FOR {$i = 1 : N; k = 1 : M_i; l = 1 : L_i; j = 1 : P$}
		\STATE compute $a_{ijkl}^{(1)}$ and $a_{ijkl}^{(0)}$ using (\ref{a_ijkl}) based on current estimate  $\bm{\Theta}_{\pi}^{(g)}$.
	\ENDFOR
	\FOR {$i = 1 : N$}
	\STATE update the current estimate of $\hat{{\pi}}_i$ using (\ref{pi_i}).
	\ENDFOR
	\FOR {$i = 1 : N, k = 1 : M_i$}
	\STATE update the current estimate of $\hat{\alpha}_{ik}$ using (\ref{alpha_ik}).
	\ENDFOR
	\FOR {$i = 1 : N, l = 1 : L_i$}
	\STATE update the current estimate of $\hat{\beta}_{il}$ using (\ref{beta_il}).
	\ENDFOR
	\STATE Update the current estimate of $\hat{\bm{\Psi}}$ by solving (\ref{EM_Mstep}).
	\STATE Update $\mbox{LLF}_{prev} \leftarrow \mbox{LLF}_{pres}$.
	\STATE Update $\bm{\Theta}_{\pi}^{(g)}$ based on the updated values of $\hat{\bm{\Psi}}, \hat{\bm{\alpha}}, \hat{\bm{\beta}}, \hat{\bm{\pi}}$.
	\STATE Update the log likelihood $\mbox{LLF}_{pres}$ using (\ref{LLF}) and the updated $\bm{\Theta}_{\pi}^{(g)}$.
	\ENDWHILE
	\FOR {$i = 1 : N$}
	\IF {$\hat{\pi}_i \ge (1 - \hat{\pi}_i)$}
	\STATE Identify the $i^{th}$ master-slave communication link as attacked.
	\ELSE
	\STATE Identify the $i^{th}$ master-slave communication link as unattacked.
	\ENDIF
	\ENDFOR
	\STATE Discard the information from master-slave communication links identified as attacked.
	\STATE Using the maximum likelihood estimates of $\hat{\bm{a}}_1$, $\hat{\bm{b}}_1$, $\hat{\bm{a}}_2$, $\hat{\bm{b}}_2$, $\hat{\bm{\alpha}}$ and $\hat{\bm{\beta}}$, compute the estimate of phase offset using (\ref{optimum_estimator}).
\end{algorithmic}
In summary, we use the EM algorithm for identifying the master-slave communication links which have been subject to delay attacks, and employ the optimal vector location parameter estimator to estimate the phase offset of the slave node.

\section{Simulation Results}\label{section4}
In this section, we evaluate the performance of the proposed POE scheme under various network scenarios. We compare the performance of the proposed POE scheme to other conventional robust POE schemes available in the literature along with the lower bounds presented in Section \ref{section3}.
\subsection{Generating the random queuing delays}\label{subsec4_1}
\begin{figure}[t]
	\centering
	\includegraphics[height = 2.0 in, width=0.75\textwidth]{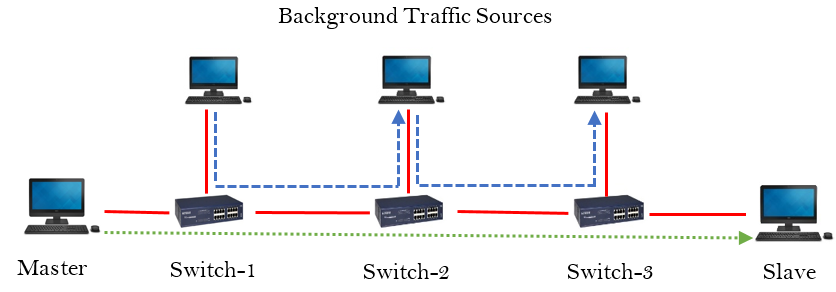}
	\caption{Example of a three switch network with cross traffic flows. The solid red lines indicate network links, dashed blue lines indicate the direction of background traffic flows, and dotted green line represents the direction of synchronization traffic flow.}\label{Traffic}
\end{figure}

\begin{figure}
	\centering
	\begin{subfigure}[b]{0.45\textwidth}
		\includegraphics[height = 2.5 in, width=\textwidth]{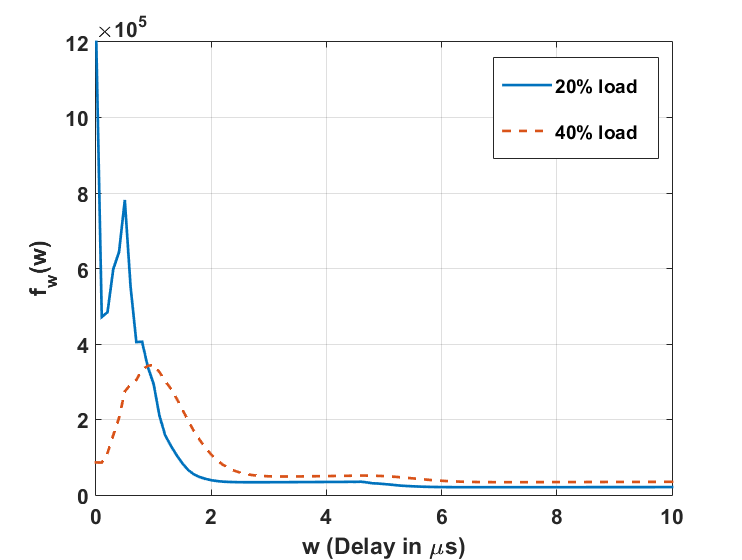}
		\caption{ }
	\end{subfigure}
	~ 
	\begin{subfigure}[b]{0.45\textwidth}
		\includegraphics[height = 2.5 in, width=\textwidth]{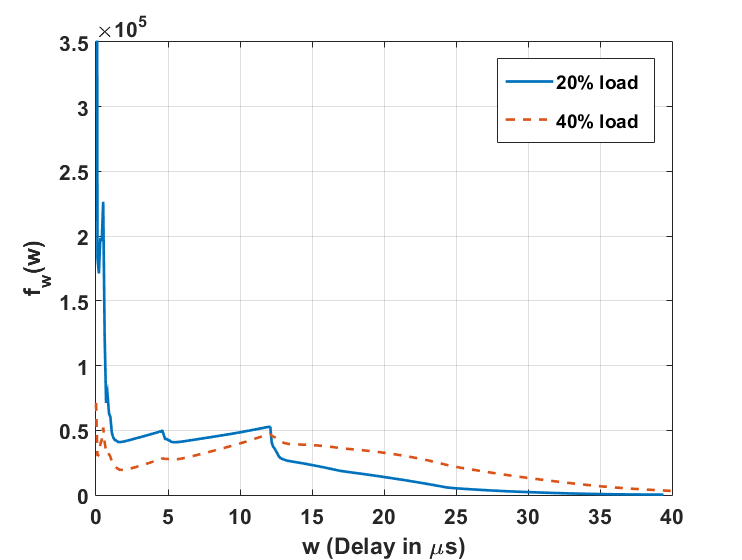}
		\caption{ }
	\end{subfigure}
	
	\begin{subfigure}[b]{0.45\textwidth}
		\includegraphics[height = 2.5 in, width=\textwidth]{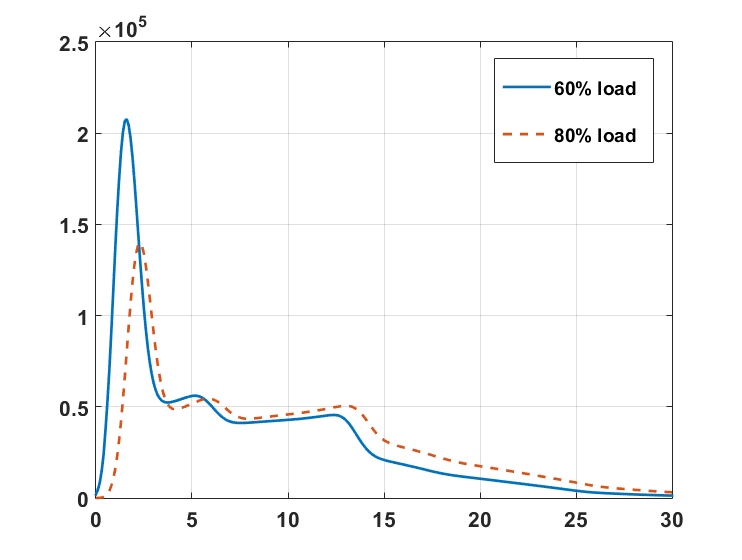}
		\caption{ }
	\end{subfigure}
	~ 
	\begin{subfigure}[b]{0.45\textwidth}
		\includegraphics[height = 2.5 in, width=\textwidth]{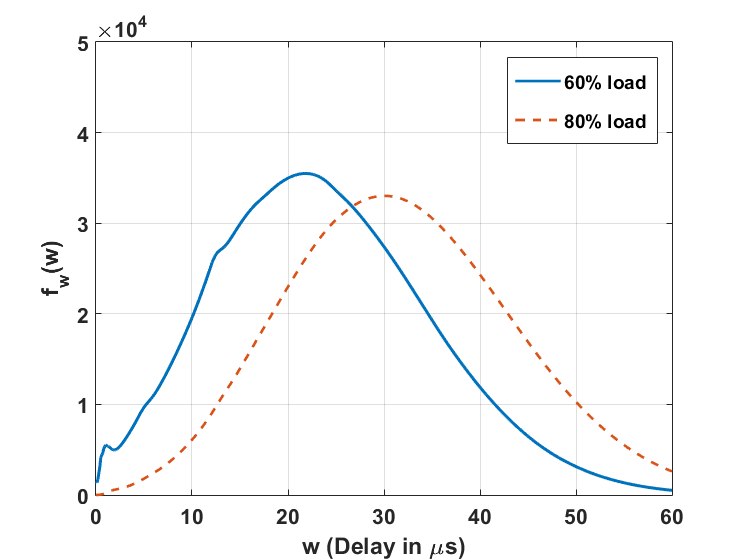}
		\caption{ }
	\end{subfigure}	
	\caption{Empirical pdf of queuing delays under various loads. (a) TM-1 network model. (b) TM-2 network model. (c) TM-1 network model. (d) TM-2 network model.}\label{f_pdf}
\end{figure}

We follow the approach given in \cite{Anand_2015}, \cite{Anand_Lest} for generating the random queuing delays in the backhaul networks. We assume a Gigabit ethernet network consisting of a cascade of $10$ switches between the master and slave node. Each switch is assumed to be a store-and-forward switch that implements strict priority queuing. We mainly consider the background cross traffic flows. In cross traffic flows, fresh background traffic is injected at each switch along the master-slave path, and this traffic exits the master-slave path at the subsequent switch [see 3-switch example in Fig. \ref{Traffic}]. The arrival times and sizes of background traffic packets injected at each switch were assumed to be statistically independent of traffic at other switches.
In the context of LTE backhaul networks, the traffic generated by other users of the network can be typically modeled as cross traffic flows. With regard to the packet size distributions of background traffic, we use Traffic Models 1 (TM1) and 2 (TM2) from the ITU-T recommendation G.8261 \cite{ITU_2008} for cross traffic flows, as described in Table \ref{sec4_table1}. 
\begin{table}
	\begin{center}
		\begin{tabular}{|c|c|c|}
			\hline
			{\bf Traf. Model} & {\bf Packet Sizes (in Bytes)} & {\bf \% of total load} \\
			\hline 
			TM-1 & \{64, 576, 1518\} & \{80\%, 5\%, 15\%\} \\
			\hline
			TM-2 & \{64, 576, 1518\} & \{30\%, 10\%, 60\%\} \\
			\hline
		\end{tabular}
		\caption{Composition of background packets in the traffic models}\label{sec4_table1}
	\end{center}
\end{table}
For the load factor, i.e., the percentage of the link capacity consumed by background traffic, we consider values between 20--80\% of the link capacity. We assume that the interarrival times between packets in all background traffic flows follow exponential distributions, and set the rate parameter of each exponential distribution to obtain the desired load factor \cite{Anand_2015}. Empirical pdfs of the queuing delays, shown in Fig. \ref{f_pdf}, were obtained using a custom MATLAB-based network simulator. Our simulations assumed that all switches were store-and-forward switches that implemented strict priority queuing. In our simulations, we consider  scenarios where the distribution of queuing delays in the forward and reverse paths are symmetrical and equal for all the master-slave communication links, i.e., $f_{1i}(w) = f_{2i}(w) = f_w(w)$.

\subsection{Approaches from the literature employed for comparison}\label{subsec4_3}
We now describe the various approaches that we compare in our simulations:
\begin{enumerate}
	\item \emph{Genie optimum approach:} In this POE scheme, we have prior knowledge of the links which have been attacked, and discard the information from these links. The phase offset $\hat{\delta}({\bf y})$ is then estimated using (\ref{optimum_estimator}). We should mention here that this approach gives a bound on the best achievable performance of a POE scheme.
	
	\item \emph{EM minimax approach-I:} In this POE scheme, we have prior knowledge of the density functions of the queuing delays. We try to identify the links which have been attacked using the EM algorithm, and fuse information from the unattacked links using (\ref{optimum_estimator}).
	
	\item \emph{EM minimax approach-II:} In this POE scheme, we try to identify the links which have been attacked as well as the mixture components of the queuing delays using the EM algorithm, and fuse  information from the unattacked links using (\ref{optimum_estimator}).
	
	\item \emph{Fault-tolerant algorithm approach:} This POE scheme was proposed in \cite{Gaderer_2010}. In this scheme, we have $prior$ knowledge regarding the number of attackers (say $M$). We first calculate the phase offset estimate for each of the individual master-slave links (using the sample mean), and we discard the $M$ lowest and $M$ largest values of the phase offset, and form the estimate of $\hat{\delta}({\bf y})$ from the remaining links deemed to be non-faulty. 
	\item \emph{Median approach}: This POE scheme was proposed in \cite{Sun_2006}.  In this scheme, we first estimate the phase offset for each of the individual master-slave links (using the sample mean). We then estimate  $\hat{\delta}({\bf y})$ as the median of estimated phase offsets from the $N$ master-slave links. 
\end{enumerate}

\subsection{Simulation results}
\begin{figure}
	\centering
	\begin{subfigure}[b]{0.4\textwidth}
		\includegraphics[height = 2.25 in, width=\textwidth]{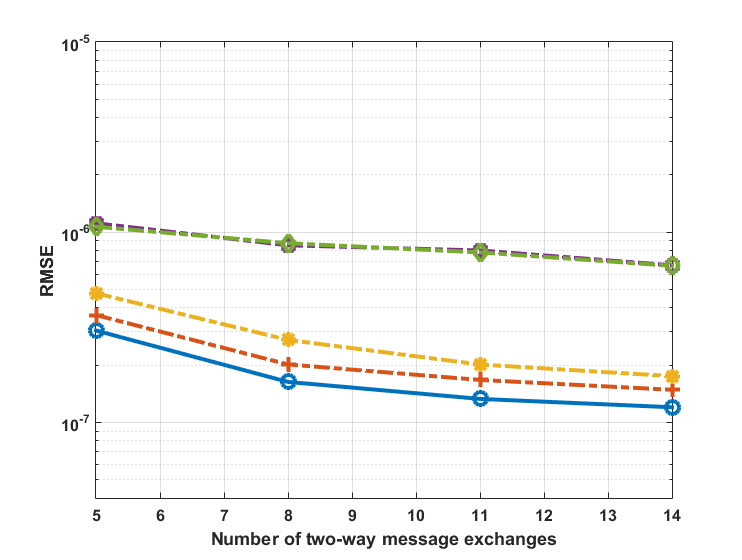}
		\caption{ }
	\end{subfigure}
	~ 
	\begin{subfigure}[b]{0.4\textwidth}
		\includegraphics[height = 2.25 in, width=\textwidth]{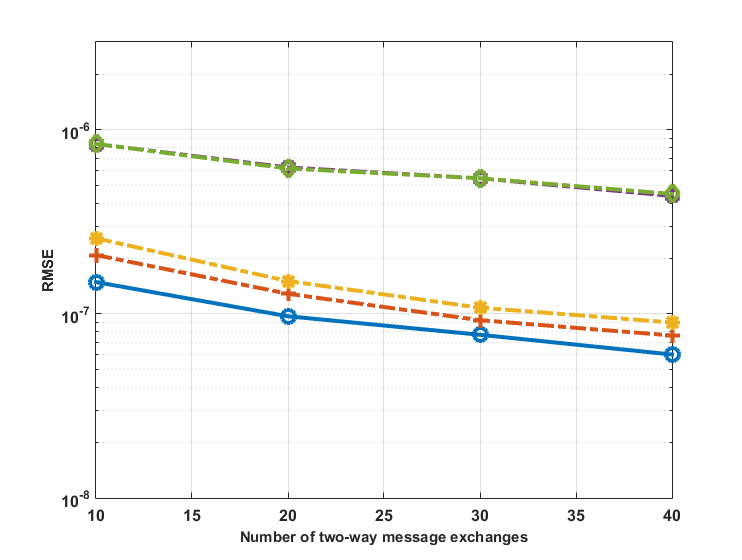}
		\caption{ }
	\end{subfigure}
	
	\begin{subfigure}[b]{0.4\textwidth}
		\includegraphics[height = 2.25 in, width=\textwidth]{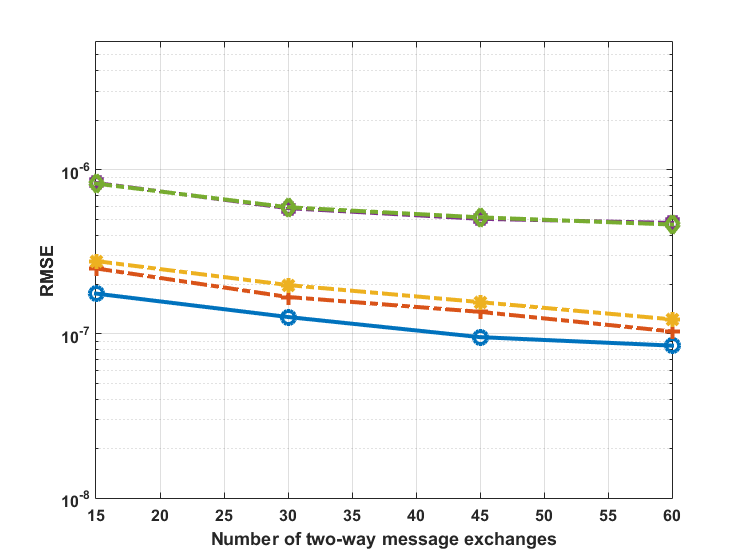}
		\caption{ }
	\end{subfigure}
	~ 
	\begin{subfigure}[b]{0.4\textwidth}
		\includegraphics[height = 2.25 in, width=\textwidth]{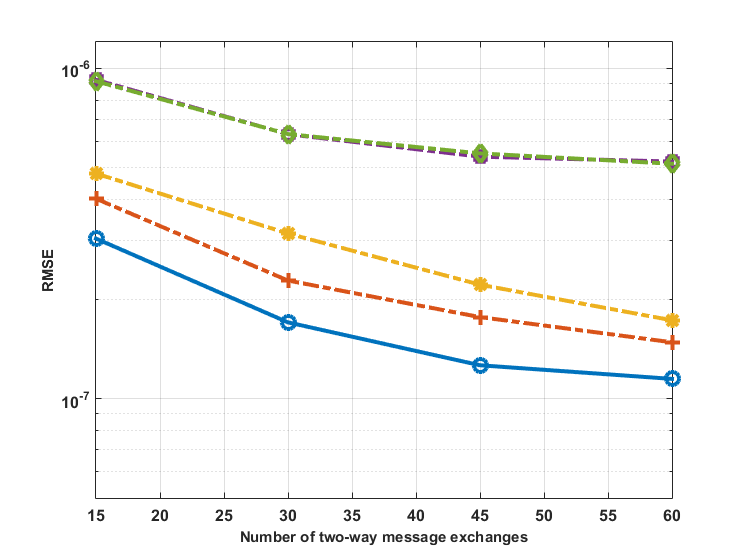}
		\caption{ }
	\end{subfigure}	
	
	\begin{subfigure}[b]{0.3\textwidth}
		\includegraphics[height = 1.2 in, width=\textwidth]{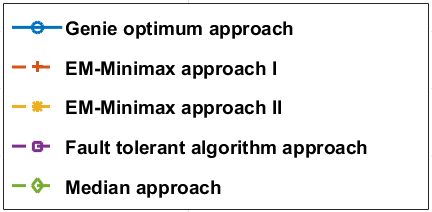}
	\end{subfigure}
	\caption{Standard deviation of estimator error with 10 switches and cross traffic flows under varying load factors in TM-1 for number of master-slave communication links = 3 and number of communication links attacked = 1. (a) 20\% load. (b) 40\% load. (c) 60\% load. (d) 80\% load.}\label{results1}
\end{figure}

\begin{figure}
	\centering
	\begin{subfigure}[b]{0.4\textwidth}
		\includegraphics[height = 2.0 in, width=\textwidth]{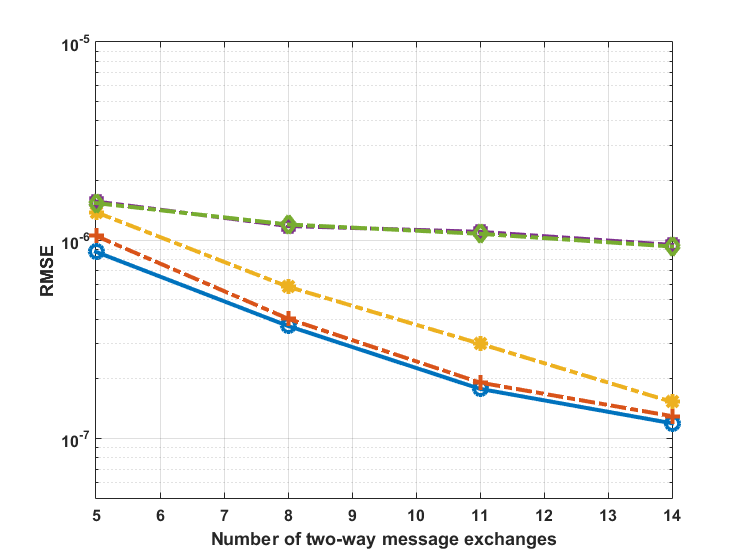}
		\caption{ }
	\end{subfigure}
	~ 
	\begin{subfigure}[b]{0.4\textwidth}
		\includegraphics[height = 2.0 in, width=\textwidth]{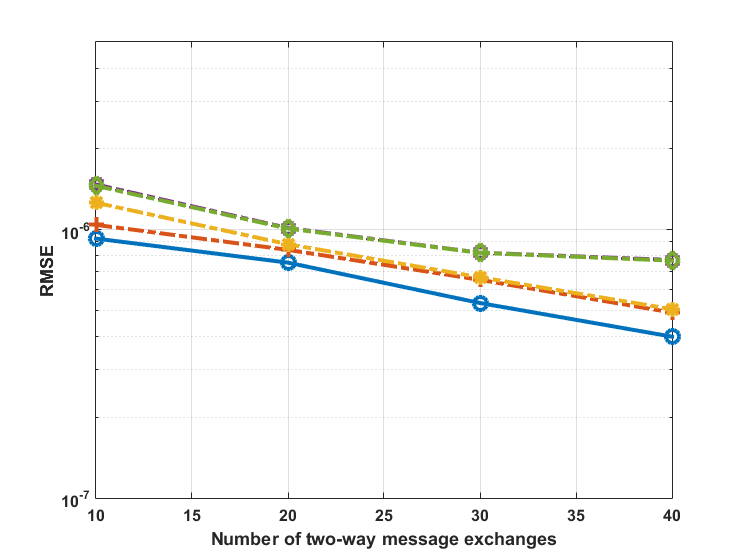}
		\caption{ }
	\end{subfigure}
	
	\begin{subfigure}[b]{0.4\textwidth}
		\includegraphics[height = 2.0 in, width=\textwidth]{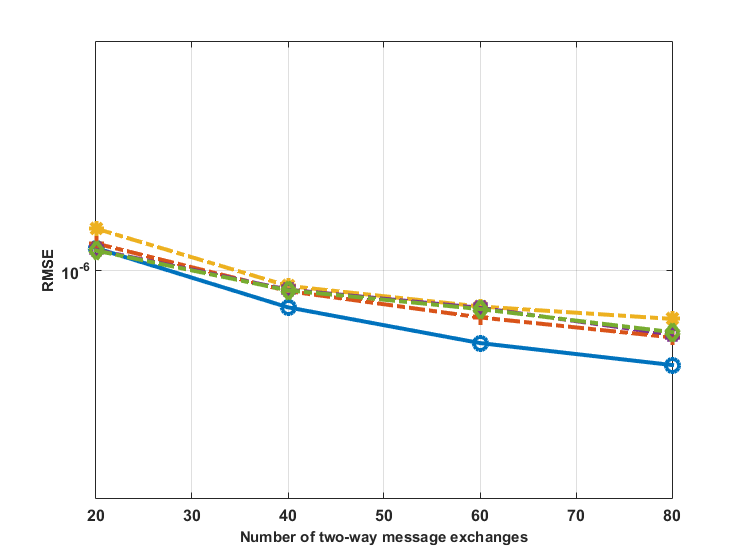}
		\caption{ }
	\end{subfigure}
	~ 
	\begin{subfigure}[b]{0.4\textwidth}
		\includegraphics[height = 2.0 in, width=\textwidth]{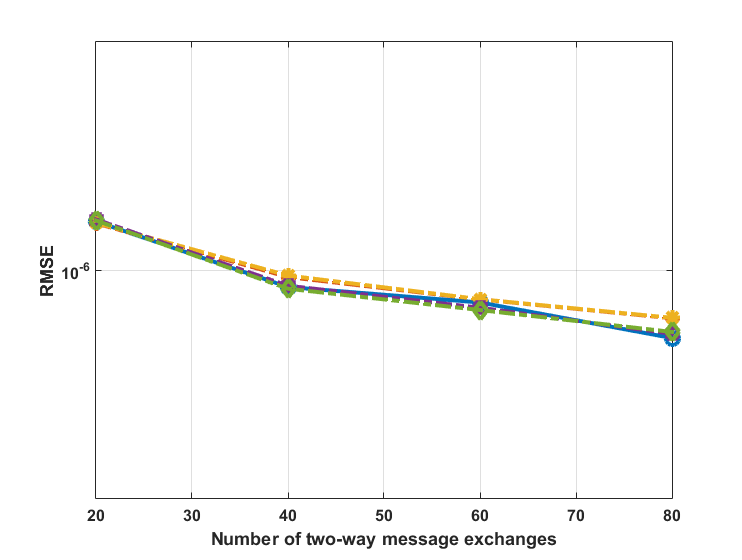}
		\caption{ }
	\end{subfigure}	
	
	\begin{subfigure}[b]{0.3\textwidth}
		\includegraphics[height = 1.2 in, width=\textwidth]{legend_comparison}
	\end{subfigure}
	\caption{Standard deviation of estimator error with 10 switches and cross traffic flows under varying load factors in TM-2 for number of master-slave communication links = 3 and number of communication links attacked = 1. (a) 20\% load. (b) 40\% load. (c) 60\% load. (d) 80\% load.}\label{results4}
\end{figure}

We carried out simulations for various network scenarios under both TM-1 and TM-2 for different loads and for various sample sizes. In our simulations, the delay attack magnitude for each attacker is chosen uniformly from the interval $[0.5, 2.0]\bigcup[-2.0, -0.5]\mbox{ }\mu s$. The results are presented in Figure \ref{results1}--\ref{results4}. In the case of 20\% load, under both TM-1 and TM-2, we approximate the queuing delay distribution by an exponential random variable (Gamma distribution with shape parameter as 1), while for the remaining cases, we approximate the queuing delay distribution by a 2-component Gamma mixture. As we can see from the results, the proposed POE scheme performs quite well under all scenarios and is relatively close to the genie optimum estimator under both network models. We briefly discuss the results:

\begin{enumerate}
	\item In Figure \ref{results1}, we study the case of $N = 3$ with one master-slave communication link being attacked under TM-1 for different loads. The proposed POE schemes perform significantly better than the median and fault tolerant algorithm (FTA) approaches under various loads. We also observe that there is no noticeable difference in performance between the median and FTA approaches. This is mainly due to the fact that in both these approaches, the smallest and largest phase offset estimates are discarded in this particular scenario. 	
	
 
	
	\item In Figure \ref{results4}, we study the case of $N = 3$ with one master-slave communication link being attacked under TM-2 for different loads. In this scenario, we see a noticeable gain in performance for the proposed POE scheme under low loads (20\% and 40\%), while at high loads all the POE schemes exhibit an estimation error close to the lower bounds. This could be due to the queuing distributions approximating a Gaussian distribution under high loads in TM-2. In such a scenario, the sample mean becomes the optimum method for fusing information from multiple master-slave communication links. 
	
		
\end{enumerate}

\section{Conclusion}\label{section5}
In this paper, we have provided a useful lower bound on the MSE for any phase offset estimation scheme in the presence of delay attacks and we show approaches that provide performance close to the bound. In particular, we presented a robust POE scheme that employs the EM algorithm along with the optimal approach for fusing information from multiple unattacked master nodes when all pdfs in the model are known. The proposed scheme does not require complete information regarding the distributions of the queuing delays, performs better than the conventional schemes available in the literature and exhibits a mean square estimation error close to the lower bounds in a number of network scenarios. Furthermore, the two-way message exchange is employed in a number of synchronization protocols including NTP \cite{NTP}, TinySeRSync \cite{Sun_2006} and the Timing-sync Protocol for Sensor Networks (TPSN) \cite{Ganeriwal_2003}. The proposed POE scheme can be easily modified for these protocols. 
\appendix[Brief discussion regarding the L-estimators]\label{Appendix1}
The POE scheme presented in \cite{Anand_Lest}, referred to as an $L$-estimator is based on a linear combination of the order statistics. The general form of these estimators for estimating the phase offset in the $i^{th}$ master-slave communication link is given by
\begin{eqnarray}\label{sec4_eq1}
\hat{\delta}_i & = & {\bf c}_1^T\tilde{\bf u}_i - {\bf c}_2^T\tilde{\bf v}_i + \eta,
\end{eqnarray}
where we have
\begin{eqnarray}
{\bf u}_i & = & [u_{i1}, u_{i2}, \cdots, u_{iP}]^T \\
{\bf v}_i & = & [v_{i1}, v_{i2}, \cdots, v_{iP}]^T \\
{\bf w}_{1i} & = & [w_{1i1}, w_{1i2}, \cdots, w_{1iP}]^T \\
{\bf w}_{2i} & = & [w_{2i1}, w_{2i2}, \cdots, w_{2iP}]^T 
\end{eqnarray}
and $\tilde{\bf u}_i$ and $\tilde{\bf v}_i$ contains all the order statistics of ${\bf u}_i$ and ${\bf v}_i$ respectively, ordered from smallest to largest and defined as follows:
\begin{eqnarray}
\tilde{\bf u}_i = \left[\min\{{\bf u}_i\}, \cdots, \max\{{\bf u}_i\}\right]^T.
\end{eqnarray}
${\bf c}_1$, ${\bf c}_2$ are weight vectors and $\eta$ is a scalar constant. Define
\begin{eqnarray}
\bm{\mu}_{ki} = E[\tilde{\bf w}_{ki}]; &  & {\bf S}_{ki} = \mbox{cov}[\tilde{\bf w}_{ki}], \\
{\bf S}_{12i} & = & E\left[\left(\tilde{\bf w}_{1i} - \mu_{1i}\right)\left(\tilde{\bf w}_{2i} - \mu_{2i}\right)\right]
\end{eqnarray}
for $k = 1, 2$. 
Let
\begin{eqnarray}
{\bf c} = \begin{bmatrix}
{\bf c}_1 \\
{\bf c}_2
\end{bmatrix}, && {\bf S}_i = \begin{bmatrix}
{\bf S}_{1i} & -{\bf S}_{12i} \\
-{\bf S}_{12i}^T & {\bf S}_{2i}
\end{bmatrix}.
\end{eqnarray}
We now present an important result from \cite{Anand_Lest} that is used for estimating the optimal values of ${\bf c}_1$ and ${\bf c}_2$ which minimize the MSE of the estimator given in (\ref{sec4_eq1}).
\begin{theorem}
	The optimum values of ${\bf c}_1$, ${\bf c}_2$ and $\eta$ that minimize the mean square error (MSE) of the estimator given in (\ref{sec4_eq1}), under the constraint of constant bias\footnote{The bias can become unbounded if not unbiased.} are
	\begin{eqnarray}
	\begin{bmatrix}
	{\bf c}_1 \\
	{\bf c}_2
	\end{bmatrix} & = & {\bf S}_i^{-1}{\bf A}^T({\bf A}{\bf S}_i^{-1}{\bf A}^T)\gamma, \\
	\eta & = & {\bf c}_1^T\bm{\mu}_{2i} - {\bf c}_2^T\bm{\mu}_{1i},
	\end{eqnarray}
	where, 
	\begin{eqnarray}
	{\bf A} = \begin{bmatrix}
	{\bf 1}_P^T & {\bf 1}_P^T \\
	{\bf 1}_P^T & -{\bf 1}_P^T 
	\end{bmatrix}, & & \gamma = [1, 0]^T,
	\end{eqnarray}
	and the resultant optimum estimator has an MSE given by
	\begin{eqnarray}\label{Lest_MSE}
	\mbox{MSE}(\hat{\delta}_i) & = & \gamma^T({\bf A}{\bf S}_i^{-1}{\bf A}^T)^{-1}\gamma.
	\end{eqnarray}
	For proof, see \cite{Anand_Lest}.
\end{theorem}

%
%
%


\begin{thebibliography}{40}
\bibitem{IEEE1588} IEEE, ``Standard for a precision clock synchronization protocol for networked measurements and control systems," IEEE 1588 (v2, 2008).
\bibitem{NTP} D. Mills, ``Network Time Protocol (Version 3) Specification, Implementation," \emph{RFC 1305}, March 1992.
\bibitem{Gaderer_2005} G. Gaderer, T. Sauter and G. Bumiller, ``Clock synchronization in powerline networks," Proceedings of International Symposium on Power Line Communications and Its Application, pp. 71--75, 2005.
\bibitem{Hadzic_2011} I. Hadzic, D. R. Morgan and Z. Sayeed, ``A synchronization algorithm for packet mans," \emph{IEEE Trans. Commun.}, vol. 59, no. 4, pp. 1142--1153, April 2011.
\bibitem{Anand_2015}A. Guruswamy, R. S. Blum, S. Kishore and M. Bordogna, ``Minimax Optimum Estimators for Phase Synchronization in IEEE 1588," \emph{IEEE Transactions on Communications}, Vol. 63 , No. 9, pp. 3350 -- 3362, Dec. 2015.
\bibitem{Ullmann_2009}M. Ullmann and M. Vogeler, ``Delay Attacks - Implication on NTP and PTP Time Synchronization," \emph{Precision Clock Synchronization for Measurement, Control and Communication, 2009. ISPCS 2009. IEEE International Symposium on}, vol., no., pp. 1--6, October 2008.
\bibitem{Gaderer_2006} G. Gaderer, A. Treytl and T. Sauter, ``Security Aspects for IEEE 1588 based Clock Synchronization Protocols," \emph{IEEE International Workshop on Factory Communication Systems}, pp. 247--250, 2006.
\bibitem{Gaderer_2008}G. Gaderer, S. Rinaldi and N. Ker$\ddot{o}$, ``Master Failures in the Precision Time Protocol," \emph{Precision Clock Synchronization for Measurement, Control and Communication, 2007. ISPCS 2008. IEEE International Symposium on}, vol., no., pp. 59--64, September 2008. 
\bibitem{Gaderer_2010}G. Gaderer, P. Loschmidt and T. Sauter, ``Improving Fault Tolerance in High-Precision Clock Synchronization," \emph{Industrial Informatics, IEEE Transactions on}, vol. 6, no. 2, pp. 206--215, May 2010.
\bibitem{Sun_2006} K. Sun, P. Ning, and C. Wang, ``TinySeRSync: Secure and resilient time synchronization in wireless sensor networks," \emph{Proceeding of the 13th ACM conference on Computer and communications security}, pp. 264-277, 2006.
\bibitem{Song_2005}H. Song, S. Zhu and G. Cao, ``Attack-resilient time synchronization for wireless sensor networks," Mobile Adhoc and Sensor Systems Conference, 2005. IEEE International Conference on , vol., no., Nov. 2005.
\bibitem{Mizrahi_2012}Mizrahi, T., ``Slave diversity: Using multiple paths to improve the accuracy of clock synchronization protocols," Precision Clock Synchronization for Measurement Control and Communication (ISPCS), 2012 International IEEE Symposium on , vol., no., pp.1--6, 2012.
\bibitem{Mizrahi_2012_2}Mizrahi, T., ``A game theoretic analysis of delay attacks against time synchronization protocols," Precision Clock Synchronization for Measurement Control and Communication (ISPCS), 2012 International IEEE Symposium on , vol., no., pp.1--6, 2012.
\bibitem{Fang_1999}Y. Fang and I. Chlamtac, ``Teletraffic analysis and mobility modeling of PCS networks," in \emph{IEEE Transactions on Communications}, vol. 47, no. 7, pp. 1062--1072, July 1999.
\bibitem{Fang_2001}Y. Fang, ``Hyper-Erlang Distribution Model and Its Application in Wireless Mobile Networks," \emph{Wireless Networks}, vol. 7, pp. 211--219, 2001.
\bibitem{Tijms_2003} H. C. Tijms, \emph{A First Course in Stochastic Models}, John Wiley \& Sons Ltd, Chichester, 2003. 
\bibitem{Zhang_2015}J. Zhang and R. S. Blum, ``Distributed joint spoofing attack identification and estimation in sensor networks," in \emph{Signal and Information Processing (ChinaSIP), 2015 IEEE China Summit and International Conference on} , vol., no., pp. 701-705, 12--15 July 2015.
\bibitem{Zhang_attacks_2015}B. Alnajjab, J. Zhang and R. S. Blum, ``Attacks on Sensor Network Parameter Estimation Systems with Quantization: Performance and Asymptotically Optimum Processing," \emph{IEEE Transactions on Signal Processing}, Vol. 63 , Issue 24, Dec. 2015, pp. 6659 - 6672.
\bibitem{Wu_2011}Y. C. Wu, Q. Chaudhrai and E. Serpedin, ``Clock Synchronization of Wireless Sensor Networks," in \emph{IEEE Signal Processing Magazine}, vol.28, no.1, pp.124--138, Jan. 2011.
\bibitem{Anand_Lest}A. Guruswamy, R. S. Blum, S. Kishore and M. Bordogna, ``On the Optimum Design of L-Estimators for Phase Offset Estimation in IEEE 1588," \emph{IEEE Transactions on Communications}, Vol. 63 , No. 9, pp. 5101 -- 5115, Dec. 2015.
\bibitem{Boyd_2004}S. P. Boyd and L. Vandenberghe, \emph{Convex optimization}, Cambridge university pressm 2004.
\bibitem{Wu_1983}C. F. J. Wu, ``On the convergence properties of the EM algorithm,"  \emph{Anna. Statist.}, vol. 14, no. 1, pp. 95--103, 1983.
\bibitem{ITU_2008} ``Timing and Synchronization Aspects in Packet Networks," \emph{Telecommunication
Standardization Sector, International Telecommunication Union (ITU), ITU-T Recommendation G.8261}, April 2008.
\bibitem{OPNET} OPNET Modeler [Online; accessed December 2013]. [Online]. Available: www.opnet.com.
\bibitem{Ganeriwal_2003}S. Ganeriwal, R. Kumar and M. B. Srivastava, ``Timing-sync protocol for sensor networks," in \emph{Proc. SenSys 03}, Los Angeles, CA, pp. 138--149, Nov. 2003.
\end{thebibliography}
\end{document}